\documentclass{emulateapj}
\usepackage{times}
\usepackage{graphicx}% Include figure files

\newcommand{\ms}

\received{October 14, 2008}
\shorttitle{$\gamma$-ray Pulsar Atlas}

\begin{document}

\title{An Atlas for Interpreting ${\gamma}$-Ray Pulsar Light Curves}

\author{Kyle P. Watters\altaffilmark{1} , Roger W. Romani\altaffilmark{1}, Patrick Weltevrede\altaffilmark{2} \& Simon Johnston\altaffilmark{2}}
\altaffiltext{1}{Department of Physics, Stanford University, Stanford, CA 94305}
\altaffiltext{2}{Australia Telescope National Facility, CSIRO, P.O. Box 76, Epping, NSW 1710, Australia}
\email{kwatters@stanford.edu}

\begin{abstract}
	
	We have simulated a population of young spin-powered pulsars and
computed the beaming pattern and light curves for the three main 
geometrical models: polar cap emission, two-pole caustic (``slot gap'')
emission and outer magnetosphere emission. The light curve shapes depend
sensitively on the magnetic inclination $\alpha$ and viewing angle
$\zeta$.  We present the results as maps of observables such as peak multiplicity 
and $\gamma$-ray peak separation in the $(\alpha, \zeta )$ plane. 
These diagrams can be used to locate allowed regions for radio-loud
and radio-quiet pulsars and to convert observed fluxes to true 
all-sky emission.
\end{abstract}

\keywords{gamma-rays - pulsars - stars: neutron}

\section{Introduction}

	With the successful launch of the {\it Fermi Gamma-ray Space
Telescope}, formerly {\it GLAST},
many pulsars are expected to be detected in the $\gamma$-ray band. This
is an important opportunity since $\sim$GeV $\gamma$-ray emission
dominates the observed electromagnetic output of young pulsars. Also,
some pulsars are expected to be detected in the $\gamma$-ray
band but undetectably faint in the radio.
The natural interpretation is
that the radio and $\gamma$-ray antenna patterns differ. Indeed, the
observed pulsars have very different radio and $\gamma$-ray pulse profiles.
Since the observed pulse profile is simply a cut through this
antenna pattern for the Earth line-of-sight angle $\zeta_E$, interpretation
of the radio and $\gamma$-ray profile requires a beaming model. When
the spin geometry is known, observed pulses can be used to select the best
model. Conversely, for a given model, the pulse pattern can restrict the
spin orientation. Finally, beaming corrections can be used to relate the
observed line-of-sight flux to the full-sky emission. Thus, aided by
beaming models, $\gamma$-ray pulsar data can give a new window into
the energetics and evolution of neutron stars in the Galaxy.

	There are three main geometrical models for the $\gamma$-ray pulse beaming,
each tied to locations in the magnetosphere where force-free
perfect MHD (${\vec E}\cdot {\vec B}=0$) conditions break down and
the uncanceled rotation-induced EMF ${\vec E}\approx r {\vec \Omega}
\times {\vec B}/c$ causes particle acceleration and radiation.
These are the so-called magnetospheric gaps.
The first $\gamma$-ray (and radio) pulsar models traced emission to
acceleration zones on the field lines 
at very low altitudes above each magnetic pole (although \citealt{dh96}
argued that such ``polar cap'' model emission could extend to  $\sim 2 R_\ast$).
Another major class of models posits ``Holloway'' gaps above the null charge surface,
extending toward the light cylinder (the ``outer gap'' model;
see \citealt{chr86}; \citealt{r96}).
More recently, it has been argued that the true null charge location depends on
currents in the gap \citep{h06}, allowing the start of the outer
gap to migrate inward towards the stellar surface.
Conversely, it has been argued that 
acceleration at the rims of the polar caps may extend to very 
high altitude \citep{mh04}, effectively pushing the polar
cap activity outward.  Together these effects suggest that emission
can occur over a large fraction of the boundary of the open zone.
One model that has been suggested 
for this geometry makes low-altitude emission visible from one
hemisphere and higher-altitude emission (i.e., above the null charge surface) visible from
the other. This is the ``two-pole caustic'' (TPC) model \citep{dr03},
which is thus intermediate between the polar cap (PC) and outer gap (OG) pictures.

\begin{figure*}[ht!]
\includegraphics[scale=0.40]{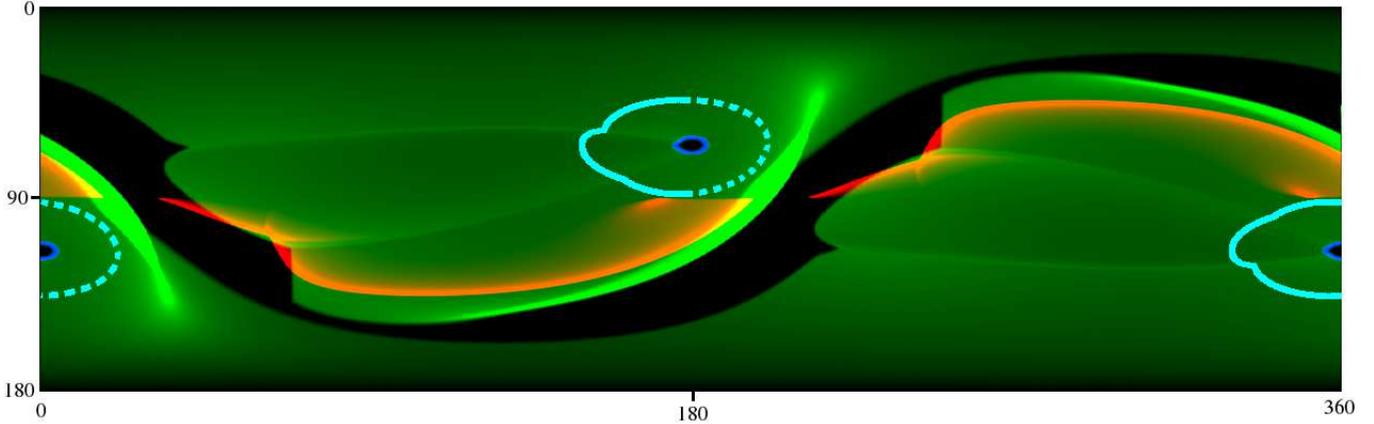}
\caption{\label{Phaseplot} Sample antenna intensity patterns for $\alpha=65\arcdeg$
with the horizontal axis of pulsar phase ($\phi$) and vertical axis of
viewing angle ($\zeta$).
The TPC model for $w=0.05$ is shown in green, the OG model for $w=0.1$ is
shown in red and the PC emission site is shown in blue. The cyan lines
show the locus of the possible high altitude ($r=500$\,km, here for $P=0.2$\,s) 
radio emission, with the radiating front half shown solid and the back half dashed
(see \S 3.3).}
\end{figure*}

\section{Simulation Method}

	We start by building a library of pulsar antenna patterns.
We follow standard practice in basing our beaming on the retarded
potential \citep{d55} magnetic dipole fields. This has the appreciable advantage
of analytic simplicity, allowing rapid computations. Numerical models of
force-free magnetospheres do exist \citep{s06}, but such models
do not have the resolution required for the sort of beaming computations pursued 
here and in any case do not, by definition, include acceleration gaps.
Still, it should be remembered that magnetospheric charges and currents
may well have an observable effect on the field structure and pulse shapes
computed here.

	We model the last closed field line surface (those field lines
tangential to the speed of light cylinder at $r_\perp=r_{LC}=cP/2\pi$), tracing these
field lines to the polar cap surface at the stellar radius $r_\ast$.
The relevant magnetospheric features in the OG and TPC models are tied to the
light cylinder and locations identified as fractions thereof; thus,
the period of the pulsar is an unnecessary parameter in mapping the structure
of the magnetosphere.
Models are computed across all possible magnetic inclinations $\alpha$.
We follow the field line structure to $<$$10^{-4} r_{LC}$,
allowing precision mapping of the PC emission region even for pulsar periods $>$1 s.
A single magnetosphere model can serve a range of pulsar periods by truncating the
grid at larger fractions of $r_{LC}$ for shorter period pulsars,
thereby maintaining the same physical distances. General
relativistic effects make small changes for $r < 3r_\ast$ but are not
included in the present sums.

	In all models, the acceleration gap is inferred to arise
near the open zone boundary tied to the last closed field lines. For
the PC picture, we simply follow radiation emitted tangentially to
these field lines at altitudes $\le$$1 R_\ast$. In the original
definition of the TPC model \citep{dr03} the emission was similarly
placed on the last closed field lines but extended to an altitude 
$r<r_{LC}$ and a perpendicular distance $r_\perp<0.75r_{LC}$, where $r_{LC}$
is the {\it perpendicular} distance to the light cylinder from the
rotation axis. In the outer gap (OG) model the emission is from
a full set of open field lines associated with the closed zone surface,
but pair production and radiation are expected to start above
the null-charge surface [${\vec \Omega} \cdot {\vec B}(R_{NC})  = 0$].

	In any physical model, however, there should be an open
zone vacuum region separating the surface of last closed field
lines from the pair formation front, which defines the radiating
surface of the gap. The OG model of \citet{r96} defined
gap thickness $w$ as the fraction of the angle from the last closed
field line to the magnetic axis that remains a charge-starved vacuum
with a large acceleration field. This characteristic gap width
$w$ is proportional to the $\gamma$-ray efficiency (see below).
A physical TPC model should also have a finite gap thickness; the
``slot gap'' model of \citet{mh03} provides a possible physical foundation
for the TPC picture.  These authors give an estimate for the
dependence of gap thickness on spin period $P$ and surface
magnetic field $B$; again the thickness increases for pulsars
putting a larger fraction of the spindown power into pulsed
$\gamma$-rays. Thus, for both models, we assume that the gap thickness
can be related to a heuristic $\gamma$-ray luminosity
%$$
%L_\gamma \approx \eta {\dot E}_{\rm SD} \approx 0.5 (\frac{{\dot E}_{\rm SD}}{10^{34}{\rm erg\ s^{-1}}})^{1/2}\times10^{34} \,\,{\rm erg\ s^{-1}}.
%\eqno (1)
%$$
$$
L_\gamma \approx \eta {\dot E}_{\rm SD} \approx C\times (\frac{{\dot E}_{\rm SD}}{10^{33}{\rm erg\ s^{-1}}})^{1/2}\times10^{33} \,\,{\rm erg\ s^{-1}}
\eqno (1)
$$
with $C$, a slowly varying function of order unity which must come from 
a detailed physical model.
This law, which assumes a $\gamma$-ray efficiency
$(w=) \eta \propto {\dot E}_{\rm SD}^{-1/2}$, is natural in models that
maintain a fixed voltage drop across the acceleration gap.
Clearly, as $w\longrightarrow 1$ this simple geometrical approximation
should break down.
The original TPC formulation has $w= 0$. In a full
physical model, the radiating pair formation front would have a
finite thickness and the radiation pattern of the particles
would have a finite width; these effects are not treated here
but would serve to smooth out the beams computed in the 
basic geometrical model.
%condition is met along a field line only after the altitude $r$
%has passed through a maximum; these cases define a topologically separate
%``high altitude gap'' -- since this may not be active in real magnetospheres
%we separately monitor radiation from these field lines. In other cases,
%the null charge condition is only met beyond the light cylinder. Such
%field lines have no outer gap at all.  
Sample
antenna patterns (sky maps of the pulsar beam) are shown for
the three models in Figure \ref{Phaseplot}.

	In each case, the antenna pattern is built up by Monte Carlo simulation.
Radiation is emitted uniformly along the gap surface. The photons are
aberrated, the time of flight to a distant observer is computed and the 
emission is assigned to a bin of pulsar phase and co-latitude
(the sky map antenna pattern; see \citealt{ry95}; \citealt{dhr04}).
We normalize this antenna pattern by summing emission at all
pulsar phases $\phi$ and along all lines-of-sight $\zeta$ and scaling by
the heuristic $\gamma$-ray luminosity  (Equation 1).

	Although we focus here on the beaming characteristics rather than the
population, to obtain a realistic sample of pulsar spin properties we 
simulate a Galactic population of young pulsars following
the method of \citet{wj08}. Pulsars are 
assigned a 
$P$ and ${\dot P}$ from the parent distribution of \citet{wj08}. 
Since we are interested in energetic pulsars, we retain objects with
$\dot E > 10^{34}$ (i.e., a max characteristic age $\tau=P/2{\dot P} \approx 10^6$\,y),
drawing until we have a sample of $\sim$175,000 pulsars. This is $\sim$12$\times$
the true Galactic density (given a birthrate of $\sim$1.5 per century); the
over-density serves to decrease statistical fluctuations. Magnetic inclinations
$\alpha$ and Earth viewing angles $\zeta$ were drawn isotropically.
Given the low upper limit on age, we ignore any possible magnetic alignment,
which, if present, appears to occur on longer time scales.  For each
drawn pulsar the appropriate $\gamma$-ray light curve is computed for the
drawn line-of-sight $\zeta$ using the appropriate antenna pattern from the 
library of sky maps.

	While we do use a realistic pulsar population, we do not report here
on predictions for the statistical distributions of $\gamma$-ray pulse shape, luminosity 
and Galactic location.
These, and predictions for the relative numbers of radio-loud and radio-quiet pulsars
expected in the LAT sky survey, are deferred to a subsequent paper. We also
do not treat here the millisecond pulsar population, as with small light
cylinder radii the difference between inner magnetosphere
and outer magnetosphere emission zones is less stark.

\begin{figure*}[h!]
\includegraphics[scale=1.35]{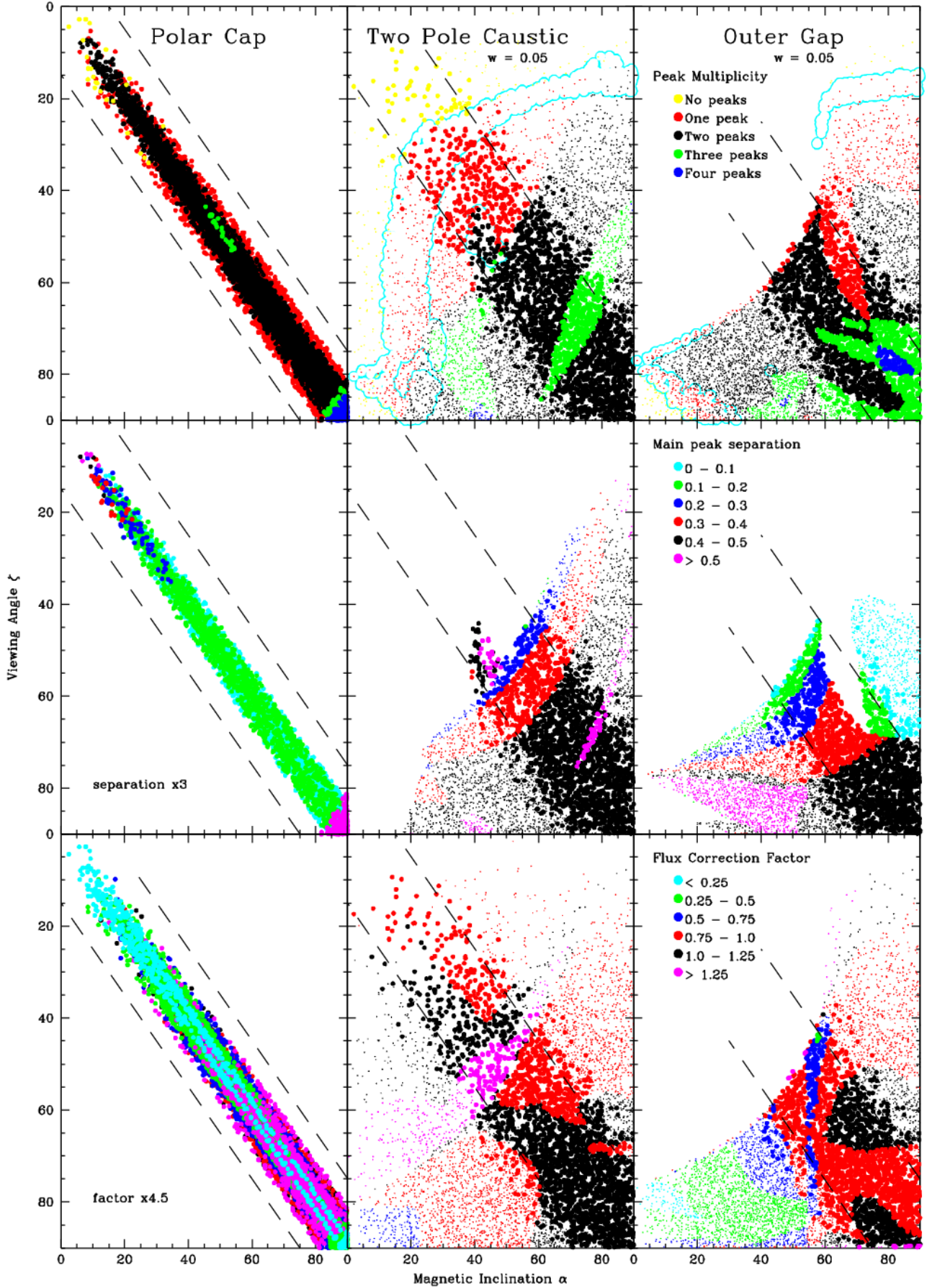}
\caption{\label{Atlas} An atlas of pulse profile properties for three
$\gamma$-ray pulsar models. Top row: pulse complexity (peak number); broad 
peaks lie in zone enclosed by the cyan lines. Middle row: pulse width $\Delta$,
the maximum separation of the major peaks. Bottom row: the 
``flux conversion factor'' $f_\Omega$ for correction to all-sky flux.
For the PC model, the small ranges of $\Delta$ and $f_\Omega$ require
re-normalization of the color scale for those panels.}
\end{figure*}

\subsection{Radio Emission}

	Radio beam models have a long history but are, if anything, 
even more heuristic than those
for the GeV $\gamma$-rays. Here we consider two scenarios. One is simple
low-altitude emission, which assumes a circular cone of radio emission along
the last closed field lines with an angular width
$$
2\rho = 10.8 P^{-1/2} \,\, {\rm degrees}
\eqno (2)
$$
centered on the magnetic axis \citep{r93,g94}.
Here, $\rho$ is the half-opening angle of the radio cone; the 
scaling is well supported by observations, which also agree
with the beam being roughly circular, as expected from dipole
field line structure at moderate altitudes.
To model this scenario in our baseline computations 
we assume radio observability whenever
the line of sight passes within $\rho$ of the magnetic axis.
The radio pulse peak is assumed to be at the same phase as this axis.

Recent evidence,
however, suggests that young, Vela-like pulsars have a fixed emission
altitude of $\sim 500-1000$\,km or $\sim 30-70 R_\ast$ \citep{kj07}.
Since these are the dominant denizens of the $\gamma$-ray pulsar
population, we also (see \S3.3) consider the effect of such wide beams
at higher altitude.
This wide beam emission is assumed to be patchy, with some lines of
sight crossing the radio zone but receiving negligible flux \citep{lm88}.
This is difficult to treat, except statistically,
and is only discussed qualitatively in this work. As we shall see in
\S3.3, it is useful to consider the possibility that the leading
portion of the radio zone has higher occupancy than the trailing half.
We thus also compute radio pulse observability for
emission only along the leading edge of a 500 km altitude radio emission zone
(Figure \ref{Phaseplot}).  Aberration and light travel effects are computed for all radio
pulses as well.

\begin{figure*}[ht!]
\includegraphics[scale=1.35]{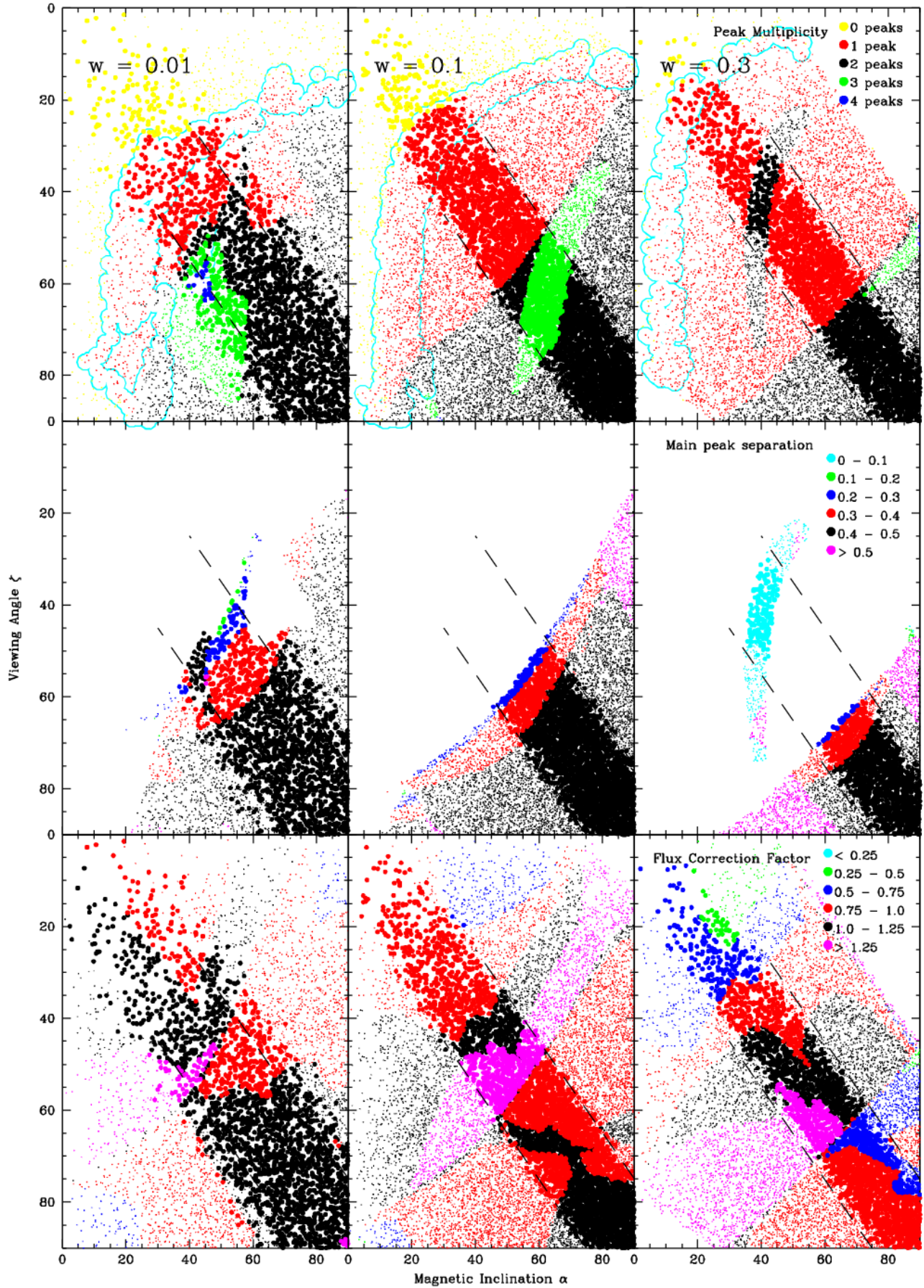}
\caption{\label{TPCAtlas} Beam shape evolution with efficiency 
for the TPC model. The three rows show peak complexity, peak
separation and flux conversion factor, as in Figure \ref{Atlas}.
The gap width (and hence $\gamma$-ray efficiency) increases left to right.}
\end{figure*}

\begin{figure*}[ht!]
\includegraphics[scale=1.35]{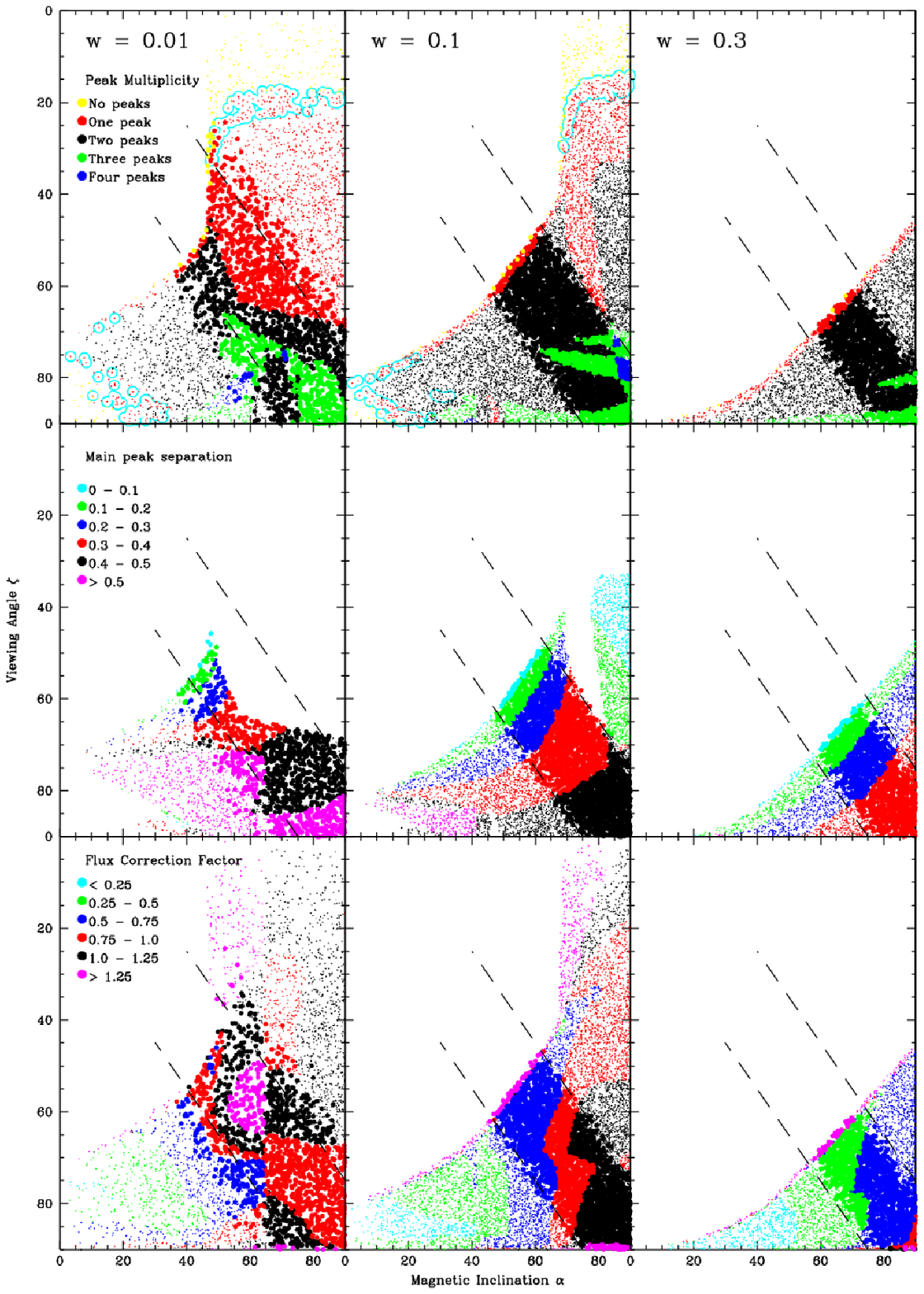}
\caption{\label{OGAtlas} Outer Gap model beam shape evolution
as a function of gap width, as in Figure \ref{TPCAtlas}.}
\end{figure*}

\subsection{Pulse Profile Characterization}

	We wish to summarize the properties of 
the $\gamma$-ray pulses of the simulated population. 
We start by flagging the major peaks for each light curve,
generated from cuts at a given $\zeta$ across antenna patterns such
as those in Figure 1. Examples of light curves for
these models can be seen in Figure 5 of \citet{dhr04} for
the particular case $\alpha=60\arcdeg$. In the Appendix
we show two summary figures of the light curves
for a range of $\alpha$ and $\zeta$ (and $w$) for the
OG and TPC models.

We have created a peak-finding routine that runs
on the $\gamma$-ray light curves produced from the models. 
The light curves are searched for maxima that
rise above the local minimum by least 20\%
of the global light curve maximum. Most non-thermal pulsar peaks are 
relatively narrow, so we flag ``sharp'' peaks by requiring 
that the the full width at half maximum (FWHM) of the 
qualifying peak (i.e. measured at 10\% of the full pulse amplitude
below the peak maximum) have $\Delta \phi <0.1$. The bright
EGRET pulsars with high S/N light curves (Crab, Vela and Geminga)
all show two peaks using this cut. However,
wider maxima may also be of interest, so we additionally flag ``broad peaks'' with
FWHM $0.1 < \Delta \phi <0.3$. These tend to appear when either
$\zeta$ or $\alpha$ is small. Note that some pulsars
with both small $\zeta$ and $\alpha$ have appreciable 
$\gamma$-ray flux but show only shallow sinusoidal variations
with no strong ``peaks'' as defined by the above algorithm.

Two additional key values regarding peak locations are also recorded.
First, we record the separation from the first peak 
%(defined as the first peak to follow the radio pulse)
to the last peak.
Additionally, we measure the lag of the first peak from the radio pulse.
To help the reader understand the results of this peak flagging,
in the Appendix figures we record the number of peaks, the broad peaks
and the pulse width (in \%) as determined by this routine for each
of the plotted light curves.

Finally, we 
compute $f_\Omega$, the ratio between the simulated total emission of the pulsar
over all observer angles $\zeta$ and the total emission if
the observed flux at $\zeta_E$ were emitted uniformly across the sky.

\section{Results}

	We have collected the geometrical pulse properties of the three
models into an ``atlas'' (Figure \ref{Atlas})
that compactly summarizes many complex pulse profiles.
We plot simulated pulsars in the ($\alpha$, $\zeta$) plane, color coding
the points to show the model properties. Small dots show objects detected in
$\gamma$-rays but not in the radio. Radio detectable objects tend to
lie at small $\beta=\zeta-\alpha$, the magnetic pole impact angle.
The dashed lines show the low-altitude standard radio
detection window for $P=0.2$\,s, the median for the observable $\gamma$-ray
pulsar population. Of course, these are not strict boundaries; pulsars
with large $P$ have smaller radio pulse widths and may 
remain radio invisible at smaller $\beta$ while especially fast pulsars
may be radio detected well beyond the plotted lines.
There are radio-detectable, $\gamma$-invisible objects as well. These 
are generally found at small $\alpha$, $\zeta$ and are not plotted here.

The top row shows the pulse complexity, the number of major pulse peaks,
with broad peaks appearing in the cyan curve.
The second row shows, for each model, the phase interval $\Delta$ between the
most widely separated major peaks.
These plots may be used to predict the $\gamma$-ray properties for a given model
and known orientation angles. Conversely, for an observed $\gamma$-ray pulse
profile one may restrict the allowed viewing angles for a given model or even rule
a model out completely. Of course, objects with well-constrained viewing geometry
{\it and} observed profiles are particularly useful as they allow the most 
strict discrimination between pulse models.
More generally, the distribution of observed pulse shapes
is different between the models. Once one assumes an underlying pulsar
distribution, such population analysis is also a very powerful discriminant
(Watters et al. 2009).

The final row of the atlas depicts a flux correction factor $f_\Omega$, useful 
in calculation of $\gamma$-ray efficiencies. The pulse profile observed along
the line of sight at $\zeta$ for a pulsar with magnetic inclination
$\alpha$ is $F_{\gamma}(\alpha, \zeta, \phi)$, which we can compute for
a given pulsar model. This means that the observed
phase-averaged flux $F_{\rm obs}$ for the Earth line-of-sight $\zeta_E$ 
is not necessarily representative of the flux averaged over the sky.
This has also been emphasized by \citet{hgg07}.
Here, we can use our model to compute the required correction 
$f_\Omega$, so that the true luminosity is
$$
L_\gamma = 4\pi f_\Omega F_{\rm obs} D^2.
\eqno (3)
$$
where $D$ is the distance to the pulsar and
$$
f_\Omega = f_\Omega(\alpha, \zeta_E) = \frac{\int\!\!\int F_{\gamma}(\alpha, \zeta, \phi)
{\rm sin}(\zeta){\rm d}\zeta{\rm d}\phi}
{2\int F_{\gamma}(\alpha, \zeta_E, \phi){\rm d}\phi}.
\eqno(4)
$$
It has become traditional to assume that the $\gamma$-ray beam
covers an area of 1 sr uniformly, so that $f_\Omega =1/(4\pi ) \approx 0.08$;
however, more modern pulsar beaming models have $f_\Omega \sim 1$ for many
viewing angles.

	The flux correction factor is crucial for estimating the
$\gamma$-ray efficiency
$$
\eta = L_\gamma /{{\dot E}_{\rm SD}} \propto f_\Omega.
\eqno (5)
$$
Thus, it appears that many recent papers have underestimated the true
$\gamma$-ray efficiency (for modern pulsar models) by an order of magnitude.

From our heuristic luminosity law, we expect that the efficiency
evolves as $\eta \propto {\dot E}_{\rm SD}^{-1/2}$. In general, one would expect
a larger fraction of the gap to remain charged-starved for more efficient
pulsars, with $w \propto \eta$.  In figures \ref{TPCAtlas} and \ref{OGAtlas}, 
we see how the pulse width and sky covering factor evolve with $w \equiv \eta $.
In particular, for low-${\dot E}_{\rm SD}$, large-$w$, high-efficiency
pulsars, we see that $f_\Omega$ decreases. Thus, the correction of 
observed phase average flux to true sky flux becomes a function 
of spindown power. As $w$ saturates at a large value of order unity,
which for this calculation is near 
${\dot E}_{\rm SD} = 10^{33} {\rm erg\ s^{-1}}$,
no radiating surfaces bound the gaps and $\gamma$-ray production ceases. 
The exact value depends on magnetic inclination $\alpha$.
\bigskip

	For the convenience of other researchers, we have collected
approximate fitting formulae for $f_\Omega$, the flux correction factor.
Because of the complexity of the beam patterns, it is always best
to use the atlas figures to determine $f_\Omega$ for a pulsar with known
parameters. However, statistical statements about luminosity and rough
population sums can benefit from analytical approximations. 
For both the TPC and OG models there is a change in the beaming
behavior at large $\alpha$ and $\zeta$. In this region one sees
emission from the hollow cone above the null charge surface that
dominates the pulsation for the classical OG and often provides the
late pulses in the TPC picture. We call this ``case I'' and thus define
an approximate boundary for such objects as
$$
\zeta > \zeta_I=(75+100w)-(60+1/w)(\alpha/90)^{2(1-w)},
\eqno (6)
$$
with $\zeta$ in degrees. Any pulsars with $\zeta<\zeta_I$ compose
case II.  For the OG scenario, $<10\%$ of the modeled
pulsars have any emission at smaller $\zeta$, so case II is not
a major contributor to the population. Such objects do, however,
exist: principally pulsars with small $w$ and large inclination $\alpha$
that are observed at small $\zeta$. The emission comes from diverging
high-altitude field lines producing faint $\gamma$-ray emission
(often with no strong pulse) over much of the sky. In the TPC
model, however, there is substantially more emission in case II,
since the flux at these angles arises from low-altitude field
lines viewed relatively near the polar cap. These light curves
tend to show shallow sinusoidal pulses. They are also missing the
caustic peaks (sharp pulses) dominating the case I objects. Thus,
in both the TPC and OG models, Equation 6 defines a boundary past which
the pulse emission is fainter and $f_\Omega$ is larger.
Figures \ref{Atlas}-\ref{AtlasKey}
and the light curves in the Appendix (Figures \ref{TPCcurves}
and \ref{OGcurves}) make these patterns clear.

	For the TPC model, we find in the case I zone
$$
f_\Omega \approx 0.8 + 1.2(0.3-w){\rm cos}(2\beta),
\eqno (7)
$$
where $\beta=\zeta-\alpha$. This applies for pulsars with $w>0.05$
and gives estimates accurate to 10\% for 76\% of the pulsars
and to 20\% for 90\% of the pulsars. In the case II zone
the flux correction factors are generally larger, with
$$
f_\Omega \approx 0.3 + 1.5(1-w)[1+(\zeta-\zeta_I)/90].
\eqno (8)
$$
This region also has more complex behavior,
so it is better to use the atlas figures if possible.
The overall accuracy for the TPC model estimates (case I
and case II zones together) is better than 10\% for 62\% of the modeled
objects and within 20\% for 82\% of all modeled pulsars.

	For the OG we consider the case I zone only, finding
$$
f_\Omega \approx 0.17 -0.69w + (1.15-1.05w)(\alpha/90)^{1.9}
\eqno (9)
$$
for $\zeta > 60\arcdeg$. A $\zeta$-dependent term appears for pulsars
viewed closer to the spin axis, so that
$$
f_\Omega \approx 0.17 -0.69w + (1.15-1.05w)(\alpha/90)^{1.9}
$$
$$
-1.35(2/3-\zeta/90)
\eqno (10)
$$
when $\zeta<60\arcdeg$.
These give approximations accurate to better than 10\% for
70\% of the simulated case I pulsars with $w>0.05$ and
are better than 20\% for 92\% of such pulsars. The atlas
figures give a guide to regions where exceptions occur.

\begin{figure*}[ht!]
\includegraphics[scale=0.93]{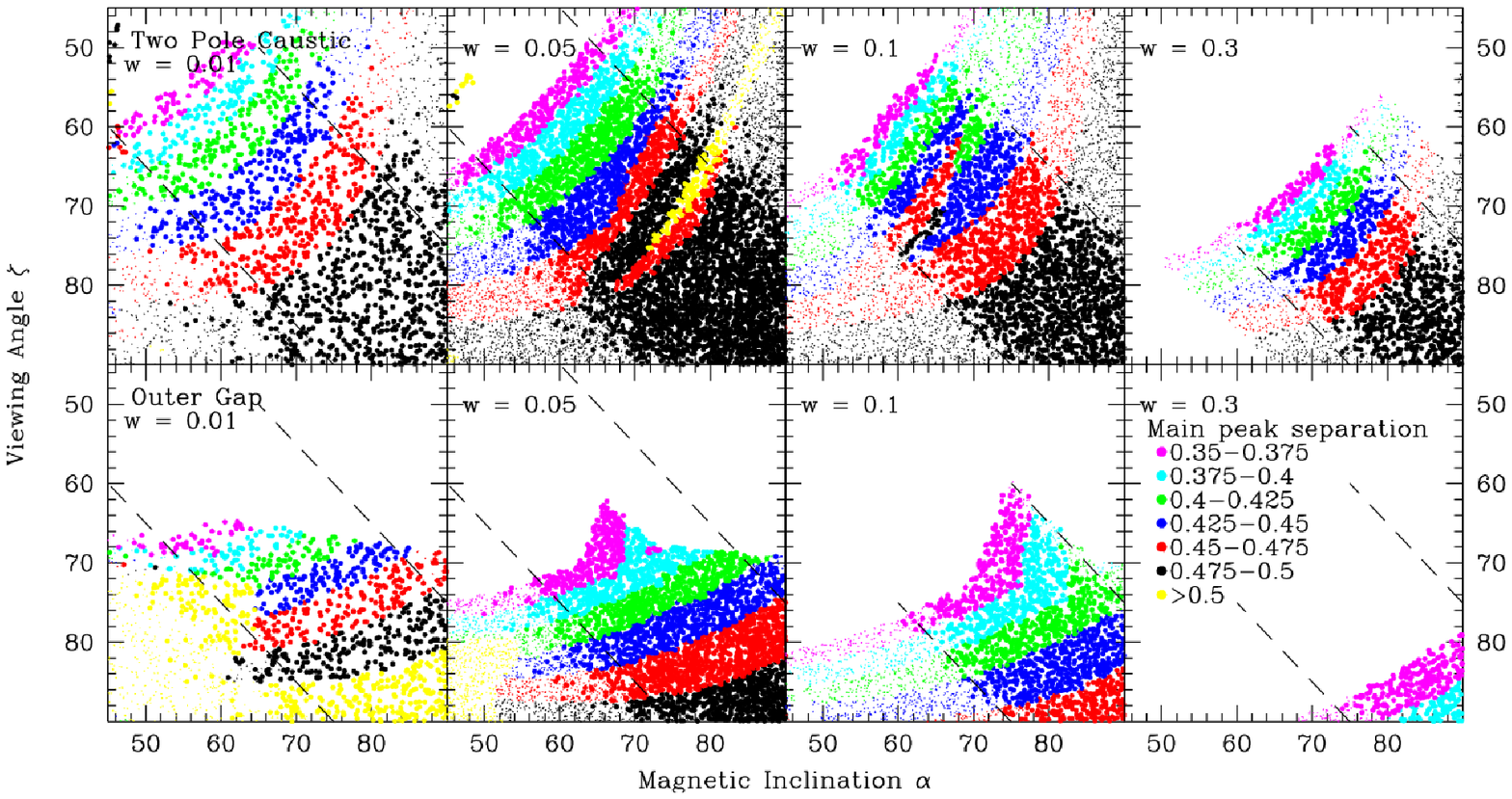}
\caption{\label{AtlasKey} The pulse width for TPC and OG models 
at several $w$. This figure should be used when possible to 
locate accurately the $\alpha$ and $\zeta$ for a given pulsar,
for reference back to Figures \ref{TPCAtlas} and  \ref{OGAtlas}.}
\end{figure*}

\subsection{Determining Individual Pulsar Geometries}

	Many young energetic pulsars produce bright, resolvable
X-ray pulsar wind nebulae (PWNe), the beam dump of relativistic
particles and fields flowing out of the magnetosphere. Often,
high-resolution X-ray images of these PWNe can be interpreted in terms
of cylindrically symmetric structures, with a mildly relativistic 
($v \approx 0.3-0.7c$) bulk outflow Doppler-boosting the intensity
along the radial flow. The inference is that the symmetry axis
traces the pulsar spin, and fits to such models \citep{nr08} can provide 
relatively model-free estimates of $\zeta_E$. Thus, for such pulsars,
the family of allowed models is given by horizontal slices across the
atlas panels.

Since we are concerned with detected $\gamma$-ray pulsars, one will
inevitably have an estimate of the pulse width
$\Delta$. This is, then, the most robust observable, and provides a
key to determining the rest of the model parameters. Thus, the
prescription for solving a pulsar for a given model is to $i$) use
${\dot E}_{\rm SD}$ (Equation 1) to estimate $w$; $ii$) go to the atlas pulse width 
plot for the closest $w$ and locate regions showing a suitable
$\Delta$; $iii$) use an external value for $\zeta_E$, when available, to
locate the relevant point in the $\Delta$ band and confirm that
that the pulsar is radio loud or quiet, as appropriate; $iv$) use the
$\zeta$ and $\alpha$ values for this point to read off other pulsar
properties (Figures 2-4). Since the $\Delta$ plot is the ``key'' to the atlas,
and since there are large regions at high $\alpha$ and $\zeta$ with
$\Delta \sim 0.4-0.5$, especially for the TPC model, we provide in
Figure 5 an expansion of the lower right quadrant of the evolving $\Delta$
panels with finer color bands. This should be used to determine
$\zeta$ and $\alpha$ when possible, which can be transferred to 
Figures 3 and 4.

	There is also a well-developed phenomenology connecting the
sweep of linear polarization within the radio pulses to cuts across
a simple magnetic pole. In the classic rotating-vector model (RVM, \citealt{rc69}),
the emission is assumed to have a polarization vector locked to the
dipole magnetic field (projected on the plane of the sky) in the emitting region.
The shape of the polarization sweep depends most sensitively on
the value sin($\alpha$)/sin($\beta$).
As $\beta$ generally takes on small values (or else the object would not be
detected in the radio in the first place),
this method tends to give poor constraints on
$\alpha$ and tighter constraints on $\beta$.
Thus, in principle, X-ray and radio data together can provide both
$\zeta$ and $\alpha$ and therefore a prediction of the $\gamma$-ray pulse shape.

In practice, the picture is often not this simple. One possible complication
arises if the altitude of the radio emission is large, where sweep-back
and time-delay can alter the naive rotating-vector model predictions \citep{ew01}.
Finally, we should reiterate that our models are purely geometrical
(albeit motivated by physics) and locked to the vacuum dipole field.
Field-line perturbations, currents and variation in emissivity along the
gap can all perturb the profiles. Nevertheless, this atlas does capture
the key differences between the models and provides a basis for broad-brush
interpretation of newly discovered pulsars.

\subsection{The Polar Cap Model}

	The polar cap picture naturally predicts that only pulsars seen at
small $\beta$ will be seen in the $\gamma$-rays. With the radio emission
originating above the surface $\gamma$-rays (which are at $r\la 2r_\ast$),
one would always expect radio emission bracketing the GeV pulse. 
``Radio-quiet'' objects, such as Geminga, are not possible in this
picture unless the radio cone is very patchy and emission is
completely absent along $\zeta_E$. ``Interpulsars,'' with emission
visible from both radio poles, are expected at the lower right 
($\alpha\sim\zeta\sim90\arcdeg$),
where the left panel of Figure 2 shows triple- and quadruple-peaked 
light curves for this model.  Note that there is also a small collection
of triply-peaked light curves at $\alpha\sim\zeta\sim45\arcdeg$ stemming from the
notch structure observed in the outline of the polar cap for inclination
angles $\alpha\sim45\arcdeg$ \citep{dhr04}.

\begin{figure}[h!]
\includegraphics[scale=0.45]{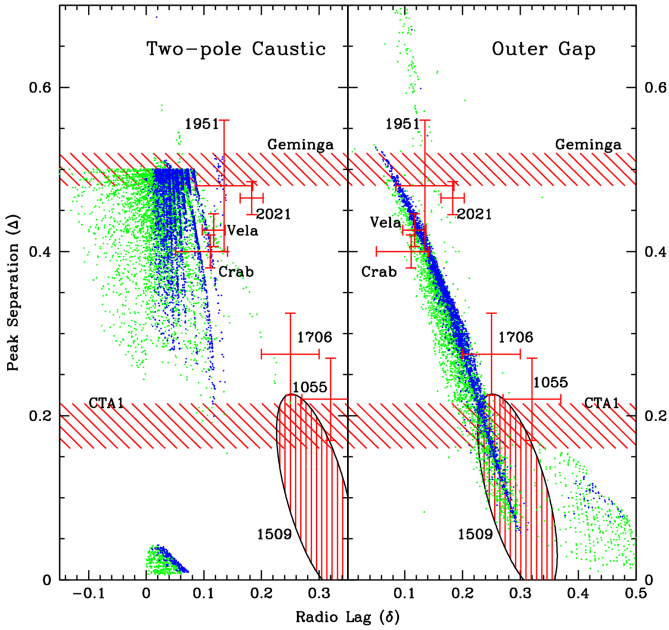}
\caption{\label{Deltadelta} Pulsar $\gamma$-ray peak separation vs. radio lag.
Green points are radio-quiet $\gamma$-ray pulsars, blue points are radio-loud
$\gamma$-ray pulsars. 
%J2021+3651 data from \citet{h08}, CTA 1 data from \citet{a08}.
} 
\end{figure}

\begin{figure}[h!]
\includegraphics[scale=0.45]{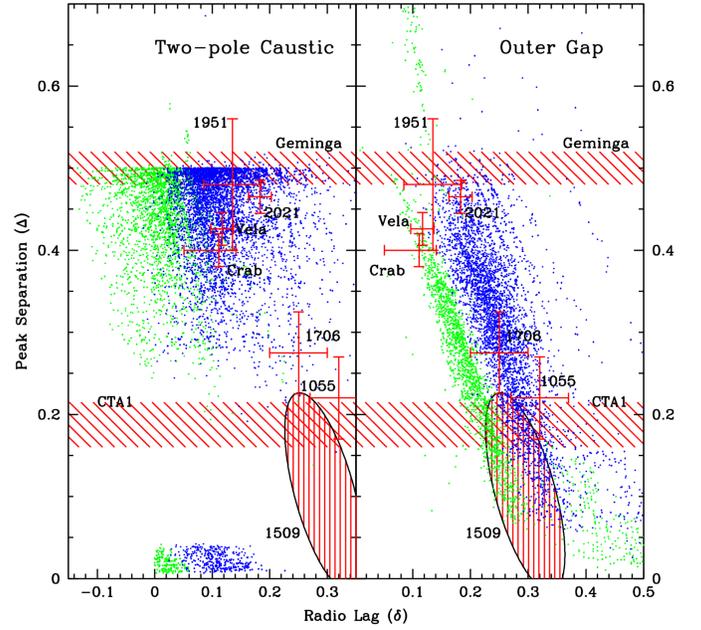}
\caption{\label{DeltadeltaHA}As Figure \ref{Deltadelta}, but assuming
high-altitude radio emission from the leading half of the radio zone.}
\end{figure}

The wide double pulses, with $\Delta=0.35-0.45$, seen for the bright
EGRET pulsars pose another challenge to the PC model.
While a double-pulse light curve is the dominant mode in the PC zone,
a wide separation can only be obtained for very small $\alpha$
and a small range of $\zeta$. Unless pulsars are born highly aligned, this
is statistically very improbable.  Additionally, model-independent measurements
of $\zeta$ made through the analysis of PWN tori mentioned above have consistently
found values $\zeta>40\arcdeg$, well beyond the range for which the polar cap model
can produce large peak separations.

These geometrical arguments, as well as considerations of flux
and spectrum,
%[Vela I reference?] 
imply that the bulk of the
observed $\gamma$-ray pulsar emission seen by EGRET could not have come from
a low-altitude cap. However, if a surface gap does have active pair
production, some $\gamma$-rays are nevertheless expected from this region; very sensitive
{\it Fermi} LAT observations may be able to uncover such a component.

\subsection{Observed Pulsars}

	The atlas lets one read off the expected $\gamma$-ray pulse properties 
for known $\alpha$ and $\zeta$ or estimate from $\Delta$ values for
undetermined $\alpha$ and $\zeta$.
It is also interesting
to compare the observed properties of specific pulsars with the models' predictions.
Table \ref{PulsarParameters} lists the observed properties of the well known EGRET
pulsars and two new detections,
J2021+3651 from {\it AGILE} \citep{h08} and CTA 1 from {\it Fermi} \citep{a08}.
Column 4 lists the
externally derived values for the line-of-sight angle $\zeta$, when available.

Note that for PSR B1055$-$52 the radio-derived $\zeta$ accords well with the value predicted by
the $\gamma$-ray pulse width $\Delta$ in the outer gap picture. The TPC picture does have
a solution matching the observed $\Delta$ (albeit only for $w$ much lower than
inferred from ${\dot E}_{\rm SD}$), but $\zeta$ does not agree with the radio
value, and additionally $\alpha$ is too small to produce the observed radio interpulse.
For the two radio-quiet pulsars Geminga and PSR J0007+7303 in CTA 1 we do not have $\zeta$
estimates. For CTA 1, both OG and TPC models can match the observed $\Delta$.
For Geminga, with $\Delta=0.5$, the TPC model has solutions along both the $\zeta$ and
$\alpha$ axes. For the OG model, Geminga-type $\Delta$ are only natural at small
$\alpha$ and large $\zeta$, where one will almost certainly miss the radio beam.

\begin{deluxetable*}{lcccccccccccc}
\tablecaption{\label{PulsarParameters} Geometry and Beaming of Bright EGRET Pulsars}
\tablehead{
  \colhead{Name} & \colhead{log(${\dot E}_{\rm SD}$)} & \colhead{w} & \colhead{d (kpc)\tablenotemark{a}}
    & \colhead{$\zeta$\tablenotemark{b}} & \colhead{$\Delta$\tablenotemark{c}}
    & \colhead{$\delta$\tablenotemark{c}} & \colhead{$\alpha_{\rm TPC}$\tablenotemark{d}} & \colhead{$\zeta_{\rm TPC}$\tablenotemark{d}}
    & \colhead{$f_\Omega$(TPC)\tablenotemark{d}} & \colhead{$\alpha_{\rm OG}$\tablenotemark{d}} & \colhead{$\zeta_{\rm OG}$\tablenotemark{d}}
    & \colhead{$f_\Omega$(OG)\tablenotemark{d}} }
\startdata
Crab         & 38.7 &0.001& $2.0\pm0.2$& 63  & 0.40  & 0.111 & 55-60 & 63    & 1.1     & 70    & 63    & 1.0\\
Vela &36.8&0.01&$0.287^{-0.017}_{+0.019}$ &64& 0.426 & 0.117 & 62-68 & 64    & 1.1-1.15& 75    & 64    & 1.0\\
B1951$+$32   & 36.6 &0.02& $2.0\pm0.6$ & ... & 0.48  & 0.135 & 55-90 & 55-90 & 1.0-1.25& 60-90 & 70-85 & 0.75-1.1\\
B1706$-$44   & 36.5 &0.02& $2.3\pm0.7$ & 53  & 0.2   & 0.25  & 45-50 & 53    & 1.05    & 47-49 & 53    & 0.7-1.0\\
J2021$+$3651 &36.5 &0.02& $3.0\pm1.0$  & 86  & 0.465 & 0.165 & 70    & 86    & 1.1     & 70    & 86    & 1.05\\
CTA 1         &35.7&0.05& $1.4\pm0.3$   & ... & 0.20  & ---   & 55-70 & 23-45 & 0.8-1.3 & 20-50 & 60-75 & 0.3-0.9\\
Geminga&34.5&0.18&$0.25^{-0.062}_{+0.120}$&...&0.50  &--- &30-80,90&90,55-80&0.7-0.9,0.6-0.8&10-25& 85    & 0.1-0.15\\
B1055$-$52   &34.5 &0.18& $0.72\pm0.15$ & 67 & 0.22  & 0.32  & 60\tablenotemark{e}&50&0.85& 78 & 67    & 0.55\\
\enddata
\tablenotetext{a}{Distance references (in order): traditional for Crab;
\citet{dod03}; 
\citet{ss00}; \citet{cl02}; \citet{vrn08}; \citet{a08}; \citet{fwa07}; \citet{cl02}.}
\tablenotetext{b}{From torus fits to X-ray images \citep{nr08} when available;
B1055$-$52 from radio RVM fit \citep{lm88}.}
\tablenotetext{c}{Derived from light curves in \citet{f95} except J2021+3651 \citep{h08} and CTA 1 \citep{a08}.}
\tablenotetext{d}{Estimated from Figures \ref{Atlas} and \ref{OGAtlas}.}
\tablenotetext{e}{Requires $w\la0.1$}
\end{deluxetable*}

In addition to the $\gamma$-ray pulse width $\Delta$, for radio-detected objects,
the lag $\delta$ of the first $\gamma$-ray pulse from the radio peak
is a second convenient observable.
In Figure \ref{Deltadelta} we have plotted in blue a sample of 
simulated $\gamma$-ray pulsars, in the $\Delta-\delta$ plane, 
assuming that the radio peak is centered along the magnetic axis. 
If the viewing angle $\zeta$ is too far from the pole (i.e., $\beta$ 
large) then no radio pulse will be detected.  In these cases we have 
plotted the pulsars in green, with $\delta$ the phase lag from the
closest approach to the radio pole. For the TPC model, many of these pulsars have
a first $\gamma$-ray peak leading the radio pole.
Some known $\gamma$-ray pulsars with well defined pulse profiles have been
placed on the figure. 
For the two ``radio-quiet'' pulsars Geminga and PSR J0007+7302, no $\delta$
is available, so the allowed regions are shown as horizontal bands.
PSR B1509-58 was seen at MeV energies by {\it CGRO}, but was not convincingly
seen by EGRET; its broad peak has a small but poorly defined $\Delta$, 
so it appears as a diagonal band.

%	The pulsars falling at small 
%$\Delta\sim0.2-0.35$ in the TPC model are produced in a manner very similar to that
%of the OG model.  The small peak separation occurs
%for moderate viewing angles for which both peaks originate 
%from the same magnetic pole -- this is exactly the ``Outer Gap'' picture.
%Of course, one can eliminate the higher altitude peak to remove
%such OG cases by tuning to a lower-altitude cut-off in the gap emission.
%Unfortunately for many angles this also eliminates flux from the second
%``TPC'' peak. Interestingly, the TPC model only manages to produce such
%separations for small $\beta$, i.e. it cannot produce radio-quiet
%objects, such as PSR J0007+7303 in CTA 1, unless some effect other than
%our modeled beaming eliminates the radio emission.

	It is clear from Figure \ref{Deltadelta} that many 
of the observed radio/$\gamma$-ray pulsars have a larger radio
lag than expected in either the TPC or OG model if the radio pulse is
centered on the magnetic axis. This suggests that some ingredient is missing
from our basic geometrical model. One possibility is that magnetospheric
currents sweep the $\gamma$-ray emitting field lines back to later
phase. Alternatively, an extra lag may be connected with the posited
high altitude radio emission of the young pulsars \citep{wj08b}.
If we invoke patchy emission and make the additional (new) assumption
that the patches are most active on the leading half of the radio zone, 
then aberration and field-line spreading on this leading half of the cone
can drive the high-altitude radio peak to much earlier phases. The
range of allowed lags is shown in Figure \ref{DeltadeltaHA} for
a fixed altitude of 500\,km. Note that the phase lead $\delta$ then depends on
the pulsar spin period, since this fixed altitude is a variable
fraction of $r_{\rm LC}$. These are the model tracks in Figure \ref{DeltadeltaHA}.
Large $\beta$ will result in a smaller increase in $\delta$
as the radio cone is cut with a small chord, further from the magnetic axis. Again for 
the green (radio-quiet pulsar) dots, we plot the $\delta$ from the
closest approach to the radio pole.  As expected, the wider radio
pulse allows fewer green (radio non-detected) objects.

	Finally, we should compare the model pulsar efficiencies 
$\eta$ with the observed $\gamma$-ray fluxes, using our computed $f_\Omega$. Of course,
this is a purely geometrical comparison,
ignoring real differences in $\gamma$-ray spectrum and the radiation
processes tapping the primary electron energy. Nevertheless, the differences
in $f_\Omega$ are substantial (and often completely ignored in discussions of
$\eta$). For gaps with near-constant brightness and pulsars with well determined
distances, the $f_\Omega$ variations can be the biggest unknown, 
making our treatment useful. 

\begin{figure}
\includegraphics[scale=0.45]{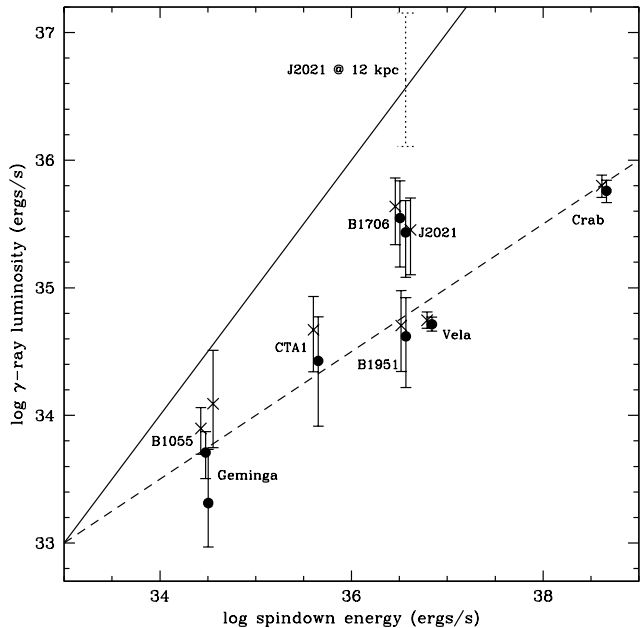}
\caption{\label{eff}Derived $\gamma$-ray luminosity, corrected 
(TPC -- crosses, OG -- circles) from the observed
phase-averaged flux to all-sky, vs. the spindown luminosity. 
Error bars include both the $d$ and $f_\Omega$ uncertainties 
(Table 1).
The dotted error bar at large $L_\gamma$ represents the $\gamma$-ray
luminosity of J2021+3651 assuming the DM distance of 12.4 kpc.}
\end{figure}

	Table \ref{PulsarParameters} contains the estimated 
$f_\Omega$ ranges inferred for the TPC and OG models. 
In Figure \ref{eff}, we plot the inferred efficiencies of the detected
$\gamma$-ray pulsars, after correction with $f_\Omega$ to simulated all-sky emission
for the two models.  The distance uncertainties generally dominate the range
in the derived $L_\gamma$.  The solid line shows complete conversion of 
spindown power to $\gamma$-rays and the dashed line shows the heuristic 
efficiency law assumed above. 

	In Figure \ref{eff}, two objects near $log({\dot E})=36.5$
deserve comment. The first is PSR J2021+3651 in the ``Dragonfly'' nebula.
This has a very large dispersion measure, and the corresponding 12.4\,kpc
distance would imply $\ge 100$\% efficient $\gamma$-ray production for
both TPC and OG pictures. However, \citet{vrn08} argue that a variety of 
pulsar and PWN measurements are more consistent with a distance of $\le 3$ kpc.
In fact, the thermal surface emission has a best fit distance of $d=2.1$\,kpc;
$d \sim 1.6$\,kpc would bring this pulsar into agreement with
the heuristic efficiency law. The origin of the large DM for this pulsar
is thus very puzzling. The next outlier is PSR B1706$-$44,
at similar ${\dot E}$, where we have adopted the DM distance of 2.3\,kpc 
\citep{cl02}. Again, smaller distances are implied by fits to the neutron
star thermal emission and the PWN energetics \citep{r05}, with preferred
values at $d\le1.8$\,kpc. However, to match the efficiency law would
require distances close to 1\,kpc; this is substantially below
the statistical lower bound on the distance estimated from H I 
absorption \citep{ket95}. Thus, the Earth line-of-sight flux of these two Vela-type
pulsars seems larger than predicted from their spindown luminosity 
and the beaming model.  Improved flux measurements from 
the LAT and improved distance estimates (e.g., a radio interferometric parallax for
PSR B1706$-$44) would be particularly valuable in tightening
up this argument.

	The most interesting region of Figure \ref{eff} is the low-${\dot E}$
range, where the modeled efficiencies approach unity. For these large
$w$ pulsars, $f_\Omega$ begins to drop significantly below unity, especially in
the OG model, decreasing the effective $L_\gamma$.  This effect dominates
as pulsars approach the $\gamma$-ray death zone, near the end of their
lives as high-energy emitters. The small $f_\Omega$ values prevent the apparent 
catastrophe of $L_\gamma>{\dot E}$ for old pulsars. Conversely, while a small fraction
of such pulsars, with their $\gamma$-beams closely confined to the spin equator,
are visible along the Earth line of sight, those that {\it are} seen have
relatively high fluxes and are detectable to relatively large distances. 
Geminga is an excellent example.

\section{Discussion \& Conclusions}

	Our study allows a quick visual summary of model predictions,
which may be compared with pulsar discoveries anticipated from the LAT.
We have already seen that the classic polar cap (PC) picture is difficult to
reconcile with the data. Both the TPC and OG pictures, however, seem
viable. Certain observations would clearly discriminate. For example,
detection of $\gamma$-ray pulsars with $\alpha+\zeta < 90\arcdeg$, but with
nearly constant intensity through the period, would give a clear preference
for the TPC model. In contrast, the prevalence of light curves with
simple double pulses separated by $0.05 < \Delta < 0.3$ is a hallmark of 
high-altitude OG emission. The distributions in peak widths and in the 
relative number of radio-loud and radio-quiet pulsars are also a strong 
discriminant; we will address these in a following paper concentrating
on the population statistics of detectable objects (Watters et al. 2009).

Also, the predicted pulse profile complexities for the two models differ.
OG models are complex for $\alpha \sim \zeta \sim 90\arcdeg$; TPC models
have complex profiles at intermediate $\alpha$. In general, the radio-faint
pulsars are double-peaked in both models, but the TPC picture has relatively
more single- and zero-peaked solutions.

One area of agreement between the TPC and OG models is the preponderance of
large $f_\Omega$; this is a principal result of our study: inferred efficiencies
should be increased by roughly an order of magnitude from those
commonly assumed.  In practice this means that some young pulsars
must have high ($>10\%$) $\gamma$-ray efficiencies. This argues that a study of
the $\gamma$-ray emission is an even better probe of magnetosphere structure
and energetics than previously thought. Our $f_\Omega$ corrections tighten 
the well-known trend for efficiency to scale as $\eta \propto {\dot E}^{-1/2}$,
although a few outliers remain.
If, as more pulsars are discovered by the LAT, we find that this
trend remains strong, we may even be able to use observed fluxes,
after $f_\Omega$ correction, for ``standard candle'' estimates of pulsar
distances. Such arguments suggest relatively low distances for 
PSRs J2021+3651 and B1706$-$44.

Importantly, for models with $w({\dot E})$ increasing with age,
$f_\Omega$ tapers off at small ${\dot E}$. Thus the sky coverage decreases
smoothly into a ``death zone'' where the gap shuts off. Near shut-off, the small solid
angle on the sky ensures that old pulsars appear relatively bright and are 
visible to large $d$.
This will be an important signature in the population sums (Watters et al. 2009).
Examination of the distribution of pulse properties over the simulated pulsar sample
is thus the key to testing the models and to producing the corrected luminosities
that will make LAT data powerful in probing pulsar populations and spindown.

\acknowledgements

	K.P.W. was supported by NASA under contract NAS5-00147. We thank the referee,
A. Harding, for several helpful suggestions that improve the utility of this paper.

%\vfill\eject
\bigskip
\bigskip
\bigskip
\bigskip
\bigskip
\bigskip
\appendix

	Here we show a selection of light curves for the two-pole caustic (TPC)
and outer gap (OG) models. For both, we arrange the sample light curves on an $\alpha$,
$\zeta$ grid so that the results may be compared with the ``atlas'' figures (Figures
\ref{Atlas}-\ref{OGAtlas}).
Light curves are shown for four $w$ values (0.01, 0.1, 0.2, 0.3). 
The number of peaks (total and broad only) and the maximum peak separation $\Delta$ (expressed here in \%)
as flagged by the automatic peak finder are given near the individual light curves.

These light curves allow the reader to visualize the topological parameters
summarized in the color ``atlas'' figures. For example, in the TPC plot one can see the broad,
sinusoidal profiles that arise at low $\alpha$ and $\zeta$ and thus earn the ``zero major
peak'' classification. One also sees how the TPC model has complex pulses near $\alpha \sim 40\arcdeg$, 
$\zeta \sim 60\arcdeg$ and that these sometimes give small separations
between the strongest peaks. For the OG plots, one sees that for $\alpha > 60\arcdeg$ and
small $\zeta$ the normal first peak is often lost; this is a result of the rapid spreading
of the high-altitude field lines contributing to this peak, which prevents formation
of the sharp leading caustic. Also, one sees that the pulse shapes are often complex
near $\zeta = 90\arcdeg$ and that the pulse shapes simplify toward basic double pulses
as $w$ becomes large.

\begin{figure*}[h!]
\includegraphics[scale=1.3]{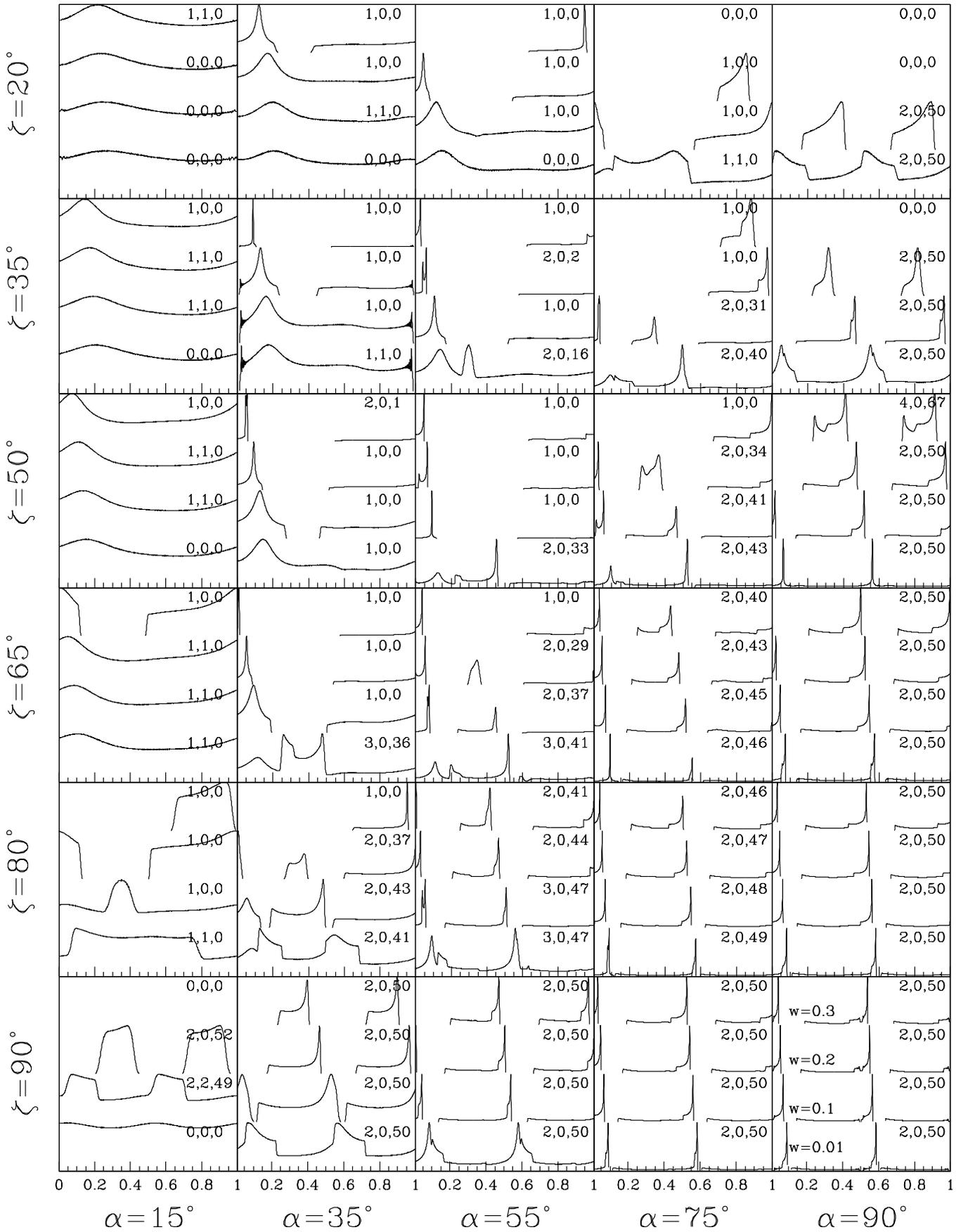}
\caption{\label{TPCcurves} Collection of sample light curves for the TPC model. 
Four select $w$ (values in bottom right panel) are shown for each panel; the radio 
pole has closest approach at phase$=$0. The values for the number of all peaks,
the number of broad peaks and the maximum peak separation (in \%) are indicated by each
curve.
Intensities are normalized to pulse maximum.}
\end{figure*}

\begin{figure*}[h!]
\includegraphics[scale=1.3]{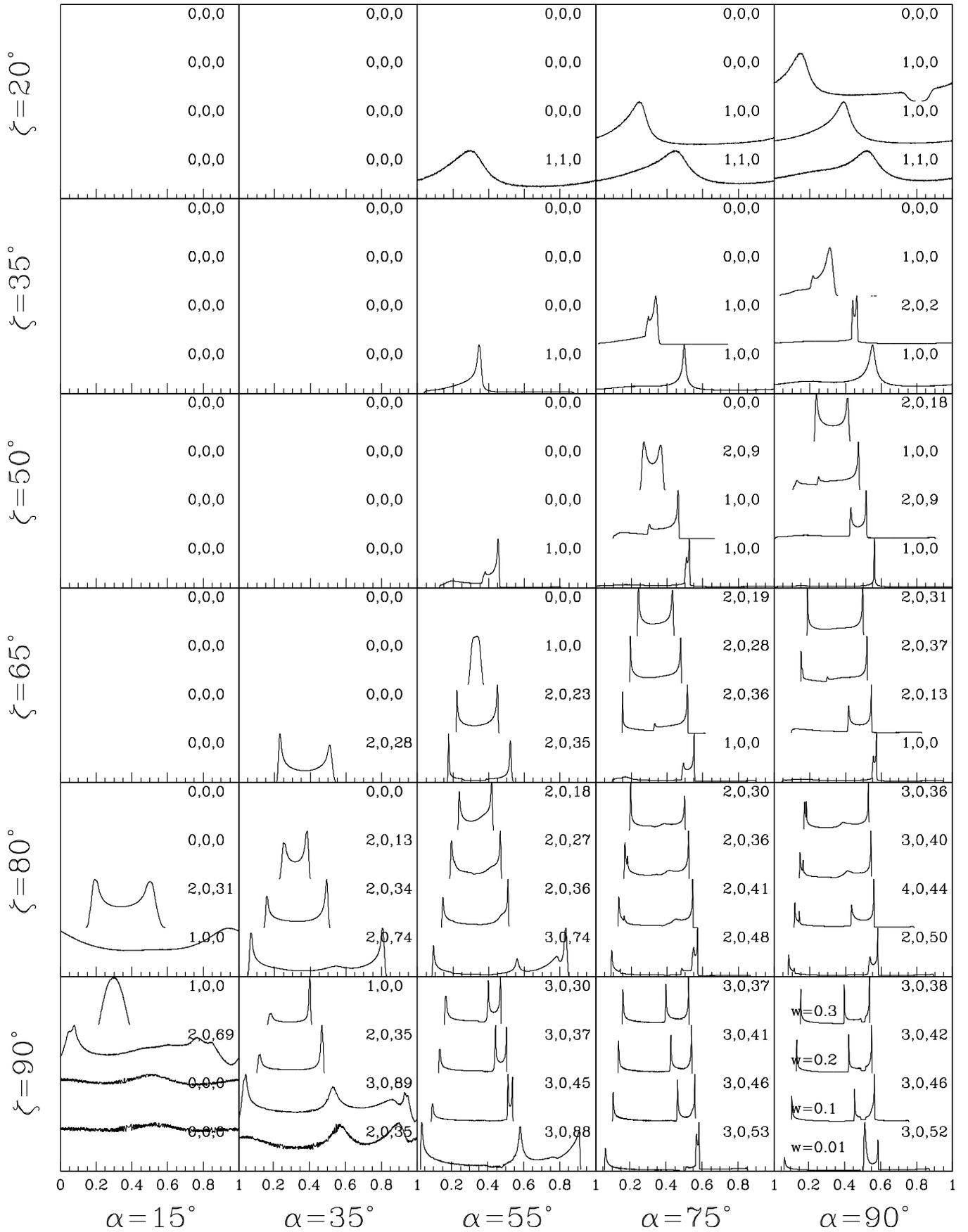}
\caption{\label{OGcurves} OG model light curves, as for Figure \ref{TPCcurves}. 
Four select $w$ (values in bottom right panel) are shown for each panel; the radio 
pole has closest approach at phase$=$0.
Intensities are normalized to pulse maximum.
For small $\alpha$ or $\zeta$, especially at large $w$, no pulses form.}
\end{figure*}


\begin{thebibliography}{}

\bibitem[Abdo et al.(2008)]{a08}Abdo, A.A. et al. 2008, Science, in press.
\bibitem[Cheng, Ho, \& Ruderman(1986)]{chr86}Cheng, K.S., Ho, C., \& Ruderman M.A. 1986, ApJ, 300, 522.
\bibitem[Cordes \& Lazio(2002)]{cl02}Cordes, J.M. \& Lazio, T.J.W. 2002, arXiv, astro-ph, 0207156.
\bibitem[Daugherty \& Harding(1996)]{dh96}Daugherty, J.K. \& Harding, A.K. 1996, ApJ, 458, 278.
\bibitem[Deutsch(1955)]{d55}Deutsch, A.J. 1955, Ann. d'Ap., 18, 1.
\bibitem[Dodson et al.(2003)] {dod03} Dodson, R., et al. 2003, \apj, 596, 1137
\bibitem[Dyks \& Rudak(2003)]{dr03}Dyks, J. \& Rudak, B. 2003, ApJ, 598, 1201.
\bibitem[Dyks, Harding, \& Rudak(2004)]{dhr04}Dyks, J., Harding, A.K., \& Rudak, B. 2004, ApJ, 606, 1125.
\bibitem[Everett \& Weisberg(2001)]{ew01}Everett, J.E. \& Weisberg, J.M. 2001, ApJ, 553, 341.
\bibitem[Faherty, Walter, \& Anderson(2007)]{fwa07}Faherty, J., Walter, F.M., \& Anderson, J. 2007, Ap\&SS, 308, 225.
\bibitem[Fierro(1995)]{f95}Fierro, J.M. 1995, PhD Thesis, Stanford University.
\bibitem[Gould(1994)]{g94}Gould, D.M. 1994, PhD Thesis, Univ. Manchester.
\bibitem[Halpern et al.(2008)]{h08}Halpern, J.P. et al. 2008, ApJ L., 688, 33.
\bibitem[Harding, Grenier \& Gonthier(2007)]{hgg07}Harding, A.K., Grenier, I.A. \& Gonthier, P.L. 2007, Ap\&SS, 309, 221
\bibitem[Hirotani(2006)]{h06}Hirotani, K. 2006, ApJ, 652, 1475.
\bibitem[Jiang \& Zhang(2006)]{jz06}Jiang, Z.J. \& Zhang, L. 2006, ApJ, 643, 1130.
\bibitem[Karastergiou \& Johnston(2007)]{kj07}Karastergiou, A. \& Johnston, S. 2007, MNRAS, 380, 1678.
\bibitem[Koribalski et al.(1995)]{ket95}Koribalski, B., et al. 1995, ApJ 441, 756.
\bibitem[Lyne \& Manchester(1988)]{lm88}Lyne, A.G. \& Manchester, R.N. 1988, MNRAS, 234, 477.
\bibitem[Manchester(2005)]{m05}Manchester, R.N. 2005, ApSS, 297, 101.
\bibitem[McLaughlin \& Cordes(2000)]{mcl00}McLaughlin, M.A. \& Cordes, J.M. 2000, ApJ, 538, 818.
\bibitem[Muslimov \& Harding(2003)]{mh03}Muslimov, A.G. \& Harding, A.K. 2003, ApJ, 588, 430.
\bibitem[Muslimov \& Harding(2004)]{mh04}Muslimov, A.G. \& Harding, A.K. 2004, ApJ, 606, 1143.
\bibitem[Ng \& Romani(2008)]{nr08}Ng, C.-Y., \& Romani, R.W. 2008, ApJ, 673, 411.
\bibitem[O'Brien et al.(2008)]{o08}O'Brien, J.T. et al. 2008, MNRAS, 388L, 1.
\bibitem[Radhakrishnan \& Cooke(1969)]{rc69}Radhakrishnan, V. \& Cooke, D.J. 1969, ApL, 3, 225.
\bibitem[Rankin(1993)]{r93}Rankin, J. 1993, ApJ, 405, 285.
\bibitem[Romani \& Yadigaroglu(1995)]{ry95}Romani, R.W. \& Yadigaroglu, I.-A. 1995, ApJ, 438, 314.
\bibitem[Romani(1996)]{r96}Romani, R.W. 1996, ApJ, 470, 469.
\bibitem[Romani et al.(2005)]{r05}Romani, R.W., Ng, C.-Y., Dodson, R., \& Brisken, W. 2005, ApJ, 631, 480.
\bibitem[Spitkovsky(2006)]{s06}Spitkovsky, A. 2006, ApJ Letters, 648, 51.
\bibitem[Story et al.(2008)]{sto08}Story, S.A., Gonthier, P.L., \& Harding, A.K. 2008, ApJ, 671, 713.
\bibitem[Strom \& Stappers(2000)]{ss00}Strom, R.G. \& Stappers, B.W. 2000, ASPC, 202, 509.
\bibitem[Van Etten, Romani, \& Ng(2008)]{vrn08}Van Etten, A., Romani, R.W., \& Ng, C.-Y. 2008, ApJ, 680, 1417.
\bibitem[Watters et al.(2009)]{wet09}Watters, K., et al 2009, ApJ, in prep.
\bibitem[Weltevrede \& Johnston(2008a)]{wj08}Weltevrede, P. \& Johnston, S. 2008, MNRAS, 387, 1755.
\bibitem[Weltevrede \& Johnston(2008b)]{wj08b}Weltevrede, P. \& Johnston, S. 2008, MNRAS, in press.
\bibitem[Yadigaroglu \& Romani(1995)]{yr95}Yadigaroglu, I.-A. \& Romani, R. 1995, ApJ, 449, 211.
\bibitem[Zhang et al.(2004)]{z04}Zhang, L., Cheng, K.S., Jiang, Z.J., \& Leung, P. 2004, ApJ, 604, 317.

\end{thebibliography}
\end{document}